\documentclass[12pt]{article}
\usepackage{graphicx}
\usepackage{longtable}

\begin{document}

\makeatletter
\renewcommand*{\@cite}[2]{{#2}}
\renewcommand*{\@biblabel}[1]{#1.\hfill}
\makeatother

\title{Photocentric orbits from a direct combination of ground-based astrometry with Hipparcos
II. Preliminary orbits for 6 astrometric binaries}
\author{G.~A.~Gontcharov\thanks{E-mail: georgegontcharov@yahoo.com}; O.~V.~Kiyaeva}

\maketitle

Pulkovo observatory, Saint-Petersburg, 196140, Russia

Key words: stars: individual (iota~Vir, zeta~UMa, gamma~UMa, kappa~Del, 20~Oph, mu~Ser)

Based on a direct combination of the Hipparcos data with astrometric
ground-based observational catalogues having epochs between 1938 and 1999
the preliminary orbits and component masses are calculated
for 6 binaries with no previous orbit calculation: 
$\iota$~Vir (HIP~69701) with period of 55 years, photocentric semi-major axis of 200 mas,
relative semi-major axis of 830 mas and a dwarf secondary of 0.6 solar masses;
$\gamma$~UMa (HIP~58001) -- 20.5 years, 90 mas, 460 mas and a dwarf secondary of 0.8 solar masses;
$\kappa$~Del (HIP~101916) -- 45 years, 100 mas, 520 mas and a dwarf secondary of 0.4 solar masses;
20~Oph (HIP~82369) -- 35.5 years, 140 mas, 460 mas and a dwarf secondary of 0.8 solar masses;
$\mu$~Ser (HIP~77516) -- 36 years, 110 mas, 350 mas and a secondary of 2.3 solar masses;
as well as a possible new component of Mizar~A ($\zeta$~UMa, HIP~65378) -- 36.5 years, 180 mas, 780 mas,
1.5 solar masses. The latter may be a pair of dwarfs.

\newpage
\section*{Introduction}

We continue the investigation of astrometric binaries among the fundamental
stars presented in the previous paper (Gontcharov and Kiyaeva
\cite[2002]{i} -- hereafter Paper I).
We use the method of a direct combination of the Hipparcos data with
astrometric ground-based observational catalogues as described by
Gontcharov et al. (\cite[2001]{pmfs}).
Briefly,
the parallaxes from the Hipparcos catalogue (ESA \cite[1997]{hip} -- hereafter HIP)
or its new reduction (van Leeuwen \cite[2007]{hip2} -- hereafter HIP2),
radial velocities from the Pulkovo Compilation of Radial Velocities
(Gontcharov \cite[2006]{rv} -- hereafter PCRV) as well as
positions from the HIP or HIP2 and observational ground-based catalogues
were used in order to reduce many observational ground-based catalogues
into a common reference frame close to the ICRS/Hipparcos.
The results of the reductions are a uniform series of
positions of 1535 Basic FK5 stars (Fricke et al. \cite[1988]{fk5b}) over several decades
(including the photocentric positions from the HIP or HIP2).
These series of star positions were used to improve individual
proper motions of the stars which were published as The Proper Motions
of Fundamental Stars catalogue (PMFS), Part I (Gontcharov et al.
\cite[2001]{pmfs}) and then slightly revised for some
stars with new ground-based and HIP2 data.
The motion of many stars appear non-linear.
It is separated into linear motion of the barycentre of the stellar pair
and elliptical motion of the photocentre around the barycentre. The latter is
analyzed here with the aim to calculate orbits and masses of components.
The procedure of separation of the non-linear motion into
a linear and an elliptic one includes a mutual improvement of the proper
motion and orbital elements in iterations as described in Paper I.

Previously we used 57 observational astrometric catalogues listed by
Gontcharov et al. (\cite[2001]{pmfs}).
Three ground-based catalogues added for the current investigation are listed
in Table~1. Mutual processing of all the 60 catalogues having epochs
between 1938 and 1999 leads to a revision of the proper motions of some stars
in the PMFS catalogue. The revised PMFS will be discussed elsewhere.
In this paper both original and revised proper motions are given.

The comparison of our results with several known orbits and application
of the method to the calculation of some new orbits were presented in the
Paper I. Preliminary orbits for another six binaries are discussed here.

\begin{table}
\def\baselinestretch{1}\normalsize\scriptsize
 \caption[]{Observational ground-based catalogues used for star positions
in addition to the list in Gontcharov et al. (\cite[2001]{pmfs})}
    \label{cats}
 \[
  \begin{tabular}{lllll}
  \hline
  Abbreviation & Observatory & Telescopes & Reference & Epoch \\
  \hline
  CDS I-144 & Herstmonceux & Cooke Transit Circle & Tucker et al. \cite[1983]{herst} & 1973 \\
  W2j00 & Washington and New Zealand & 6- and 7-inch Transit Circles  & Rafferty and Holdenried \cite[2002]{w2j00} & 1991 \\
  HAMC  & El Leoncito    & San Fernando Automatic MC & HAMC CD \cite[2001]{hamc} & 1998 \\
  \hline
  \end{tabular}
 \]
\end{table}

\section*{Stars under consideration}

The stars are listed in Table~2.
The fainter component is unseen in all the cases except $\mu$~Ser.
The rotational velocities, Teff, Fe/H, ages, masses of
the observable components and their precisions are taken from the precise multiband photometry, spectroscopy or
interferometry by
Nordstrom et al. (\cite[2004]{gcs}), Holmberg et al. (\cite[2009]{gcsiii}),
Soubiran et al. (\cite[2008]{soubiran}),
Takeda (\cite[2007]{takeda}), Feltzing and Gonzalez (\cite[2001]{feltzing}), Morossi et al. (\cite[2002]{morossi}),
Abt and Morrell (\cite[1995]{abtmorrel}), Hummel et al. (\cite[1998]{hummel}), Balachandran (\cite[1990]{bala}),
Chereul et al. (\cite[1999]{ccb}).

Table~3 gives the
components ($\mu_{\alpha}\cdot \cos\delta$ and $\mu_{\delta}$) of their
proper motion taken from the FK5, FK6 (long-term prediction (LTP) by
Wielen et al. \cite[1999]{fk6}), HIP, HIP2, PMFS catalogue and its current revision.
The PMFS and its revision give the barycentric proper motions.
The FK5 and FK6 give the photocentric proper motions based on more than
a century of observations.
These motions are close to the barycentric ones for many stars.
The HIP and HIP2 give the photocentric proper motion of these
binaries in the course of the mission.
Three years of the mission is much less than orbital periods of these stars.
Therefore the HIP/HIP2 photocentric proper motion generally is far from
the barycentric one.

\begin{table}
\def\baselinestretch{1}\normalsize\scriptsize
\caption[]{Stars under consideration: names and numbers from various catalogues, coordinates,
spectrum, visual magnitude, B-V, parallax from HIP and HIP2 with its precision,
type of HIP2 component solution, radial velocity with its precision from PCRV
and the rest parameters taken from Nordstrom et al. (\cite[2004]{gcs}), Holmberg et al. (\cite[2009]{gcsiii}),
Soubiran et al. (\cite[2008]{soubiran}),
Takeda (\cite[2007]{takeda}), Feltzing and Gonzalez (\cite[2001]{feltzing}), Morossi et al. (\cite[2002]{morossi}),
Abt and Morrell (\cite[1995]{abtmorrel}), Hummel et al. (\cite[1998]{hummel}), Balachandran (\cite[1990]{bala}),
Chereul et al. (\cite[1999]{ccb}).}
    \label{tab1}
 \[
  \begin{tabular}{lcccccc}
  \hline
  \noalign{\smallskip}
 & $\iota$~Vir & $\zeta$~UMa & $\gamma$~UMa & $\kappa$~Del & 20~Oph & $\mu$~Ser \\
  \hline
  \noalign{\smallskip}
Name                  &                  &  Mizar~A      & Phecda        &                 &                  &              \\
HIP                   &   69701          &  65378        & 58001         & 101916          & 82369            & 77516        \\
FK5                   &   525            & 497           & 447           & 772             & 1438             & 585          \\
HR                    &   5338           & 5054          & 4554          & 7896            & 6243             & 5881         \\
HD                    & 124850           & 116656        & 103287        & 196755          & 151769           & 141513       \\
ADS                   &                  & 8891          &               & 14101           &                  &              \\
$\alpha, \circ$ J2000 & 214.0036         & 200.9814      & 178.4577      & 309.7824        & 252.4585         & 237.4050     \\
$\delta, \circ$ J2000 &  -6.0005         &  54.9254      &  53.6948      &  10.0862        & -10.7830         &  -3.4302     \\
Spectrum              & F7IV-V           & A2V+A2V       & A0V           & G1IV            & F7IV             &  A0V         \\
V mag                 & 4.1              & 2.2           & 2.4           & 5.1             & 4.6              & 3.7          \\
B-V                   & 0.52             & 0.05          & 0.04          & 0.68            & 0.49             & -0.02        \\
HIP $\pi$, mas        & 47               & 42            & 39            & 33              & 27               & 21           \\
HIP2 $\pi$, mas       & 45.0$\pm$0.2     & 38.0$\pm$1.7  & 39.2$\pm$0.4  & 33.2$\pm$0.3    & 31.3$\pm$2.2     & 19.2$\pm$0.4 \\
HIP2 solution         & 5-parameter      & 5-parameter   & stochastic    & 5-parameter     & stochastic       & 7-parameter  \\
V$_{r}$, km/s         & 12.4$\pm$0.8     & -6.9$\pm$1.0  & -11.9$\pm$1.1 & -53.5$\pm$0.7   & -1.6$\pm$1.3     & -9.4$\pm$2.7 \\
$v\sin~i$, km/s       & 16               & 25            & 165           & 4               & 11               &  85          \\
$\lg$(Teff)           & 3.79             &  3.95         &               & 3.75            & 3.80             &              \\
Fe/H                  & $-0.31$ to $-0.01$ &               &               & $0.00$ to $+0.09$ & $-0.08$ to $-0.01$ &              \\
Age, Gyr              & 1.9$\div$2.1     & 0.3           & 0.3           & 2.8$\div$3.0    & 1.6$\div$1.8     &              \\
Mass, M$_{\odot}$     & 1.5$\pm$0.05     & 4.9$\pm$0.1   &               & 1.45$\pm$0.05   & 1.72$\pm$0.1     &              \\
  \hline
  \end{tabular}
 \]
\end{table}

\begin{table}
\def\baselinestretch{1}\normalsize\scriptsize
\caption[]{The proper motion of the binaries (in mas/yr):
from the FK5, FK6 (long-term prediction), HIP, HIP2, PMFS and revised PMFS}
    \label{pm}
 \[
  \begin{tabular}{lcccccc}
  \hline
  \noalign{\smallskip}
   Name &      FK5   &   FK6    &      HIP    & HIP2      & PMFS & revised PMFS \\
  \hline
  \noalign{\smallskip}
        & $\mu_{\alpha}\cdot \cos\delta~~~~~~\mu_{\delta}$ & $\mu_{\alpha}\cdot \cos\delta~~~~~\mu_{\delta}$
        & $\mu_{\alpha}\cdot \cos\delta~~~~~~\mu_{\delta}$ & $\mu_{\alpha}\cdot \cos\delta~~~~~~\mu_{\delta}$
        & $\mu_{\alpha}\cdot \cos\delta~~~~~~\mu_{\delta}$ & $\mu_{\alpha}\cdot \cos\delta~~~~~~\mu_{\delta}$ \\
  \hline
  \noalign{\smallskip}
$\iota$~Vir  & $-4~-432$  & $-11~-433$  & $-26~-420$  & -26~-419  & $-21~-433$ & $-16~-434$ \\
$\zeta$~UMa  & $+122~-20$ &             & $+121~-22$  & +119~-26  & $+121~-23$ & $+118~-22$ \\
$\gamma$~UMa & $+95~+12$  & $+95~+10$   & $+108~+11$  & +108~+11  & $+96~+8$   & $+94~+8$ \\
$\kappa$~Del & $+321~+22$ &             & $+324~+21$  & +324~+22  & $+320~+22$ & $+318~+22$ \\
20~Oph       & $+96~-93$  &             & $+94~-82$   & +72~-79   & $+95~-94$  & $+94~-92$ \\
$\mu$~Ser    & $-86~-24$  &             & $-98~-27$   & -100~-26  & $-88~-27$  & $-89~-25$ \\
  \hline
  \end{tabular}
 \]
\end{table}

In Table~4 we show the semi-major axis of apparent ellipse $a_{app}$,
standard deviation of the astrometric observations from the best orbit along $\alpha$ --
$\sigma_{\alpha}^{min}$ (in milliarcseconds, hereafter mas),
the same along $\delta$ -- $\sigma_{\delta}^{min}$,
the same along celestial great circle -- $\sigma_{\alpha+\delta}^{min}$,
and the signal-to-noise ratio (S/N) calculated as $2a_{app}/\sigma_{\alpha+\delta}^{min}$.
The standard deviations give an estimate of the accuracy of the formed
series of astrometric observations.

The deviation of the observations from the orbits shows that the used ground-based
catalogues can be divided into two groups depending
on the accidental accuracy (after the reduction to the ICRS/Hipparcos):
the classical catalogues with accuracy of positions near 100$\div$200 mas
and the catalogues obtained with the automated photoelectric telescopes of the 1980-90ths
such as Carlsberg and Bordeaux meridian circles with accuracy of positions near 30$\div$50 mas.
To calculate the best orbits the positions are weighted according to their conventional accuracy:
100 mas for the classical catalogues, standard deviation from the HIP/HIP2 - for the photoelectric catalogues
and 10 mas - for the HIP2 results (except Mizar~A for which no Hipparcos position is used).
The HIP2 results are presented as three positions for the epochs
1990.25, 1991.25 and 1992.25 calculated from the HIP2 according to usual 5-parameter solutions for
$\iota$~Vir and $\kappa$~Del, stochastic solutions for Phecda and 20~Oph and 7-parameter
(acceleration) solution for $\mu$~Ser.

It is evident from the standard deviation of the observations and the precision of the
orbital elements that our orbits must be regarded as preliminary.
They are meant to provide a baseline for future observations and
orbit calculations.

In order to evaluate the precision of the orbits a dense set of the orbits is calculated for the 7-dimension
space determined by the orbital elements.
The precision of every orbital element is calculated as the dispersion of the element for all the orbits with
$\sigma_{\alpha+\delta}<\sigma_{\alpha+\delta}^{min}+\sigma_{\alpha+\delta}^{min}/N^{1/2}$,
where N is the number of used catalogues.

The self-consistent sets of orbital elements with their precision as well as other parameters of the
binaries are presented in Table~5. To obtain the sets
we used the following common relations as mentioned in Paper I.
The orbital elements P, T, i, e and $\Omega$ are the same for the relative
(secondary with respect to primary) and photocentric
(photocentre with respect to barycentre) orbits; $\omega$ differs by $180^{\circ}$.
The ratio of the distance `barycentre -- primary' to `secondary -- primary'
is $B=M_{B}/(M_{A}+M_{B})$,
where $M_{A}$ and $M_{B}$ are the component masses; the ratio of the distance
`photocentre -- primary' to `B--A' is $\beta=1/(1+10^{0.4\Delta m})$, where $\Delta$m
is the magnitude difference. Thus, $a_{pm}=a_{BA}\cdot(B-\beta)$, where
$a_{BA}$ and $a_{pm}$ are the relative and photocentric semi-major axes
(both in milliarcseconds (mas) hereafter).
For every pair except $\mu$~Ser the mass of the primary is approximately estimated
from the magnitude, spectrum, color index or other external data.
Then the system of 2 equations
$$
\begin{array}{c}
M_{A}+M_{B}=a_{BA}^{3}/(\pi^{3}P^{2})\mbox{, ($\pi$=parallax, P=period),}\\
a_{pm}=a_{BA}\cdot(M_{B}/(M_{A}+M_{B})-1/(1+10^{0.4\Delta m}))
\end{array}
$$
is solved for $M_{B}$ and $a_{BA}$. 
Hereafter all the masses are given in solar masses.
The speckle-interferometric observations of the secondary of $\mu$~Ser
allow us to calculate $a_{BA}$ and solve these equations for $M_{A}$ and $M_{B}$.

The precisions of the orbital elements, parallax, M$_{A}$ and $\Delta m$ determine
the ones of the derived parameters. Both the accepted and derived precisions are
indicated in Table~5,
except some parameters of Phecda because of the uncertainty of its mass.

Our observational material together with the photocentric orbits is shown
in Fig.~1--6, each of which is composed of three subfigures.
The first subfigure shows the observations together
with the orbit in the form of the offsets $\Delta\delta$ versus $\Delta\alpha \cos\delta$
given in arcsec. The Hipparcos position matches the calculated orbit quite well.
The barycentre is marked by a cross. The observations are connected by O-C
(observed - calculated) lines to the appropriate epochs on the orbital ellipses.
The second and third subfigures give the offsets $\Delta\alpha \cos\delta$
and $\Delta\delta$, respectively, as a function of time.
The figures are not ideal representations of our results
because a) the astrometric catalogues have different accuracies and respective
weights in our processing and b) some astrometric catalogues containing
only $\alpha$ or $\delta$ are not presented in the subfigure `$\Delta\delta$
versus $\Delta\alpha \cos\delta$'.

\begin{table}
\def\baselinestretch{1}\normalsize\small
\caption[]{The semi-major axis of apparent ellipse $a_{app}$,
standard deviation of the astrometric observations
from the best orbit along $\alpha$ -- $\sigma_{\alpha}^{min}$ (in mas),
the same along $\delta$ -- $\sigma_{\delta}^{min}$,
the same along celestial great circle -- $\sigma_{\alpha+\delta}^{min}$,
and the signal-to-noise ratio (S/N)}
    \label{accuracy}
 \[
  \begin{tabular}{lrrrrr}
  \hline
  \noalign{\smallskip}
  Name & $a_{app}$ & $\sigma_{\alpha}^{min}$ & $\sigma_{\delta}^{min}$ & $\sigma_{\alpha+\delta}^{min}$ &  S/N \\
  \hline
  \noalign{\smallskip}
$\iota$~Vir  & 200 & 42 & 51 & 70 & 5.7 \\
$\zeta$~UMa  & 180 & 34 & 56 & 68 & 5.3 \\
$\gamma$~UMa &  90 & 19 & 23 & 33 & 5.5 \\
$\kappa$~Del & 100 & 31 & 30 & 49 & 4.1 \\
20~Oph       & 140 & 39 & 29 & 52 & 5.4 \\
$\mu$~Ser    & 100 & 37 & 29 & 51 & 3.9 \\
  \hline
  \end{tabular}
 \]
\end{table}

\begin{table}
\def\baselinestretch{1}\normalsize\scriptsize
\caption[]{Self-consistent sets of some parameters of the binaries:
7 rows of assumed (in italic) and derived orbital elements with precision
followed by the rows of other parameters where parallax, $\Delta m$,
as well as mass (except $\mu$~Ser), spectrum and V magnitude of the primary are accepted.
}
    \label{sets}
 \[
  \begin{tabular}{lllllll}
  \hline
  \noalign{\smallskip}
  Parameter            & $\iota$~Vir    & $\zeta$~UMa     & $\gamma$~UMa      & $\kappa$~Del      & 20~Oph         & $\mu$~Ser       \\
  \hline
  \noalign{\smallskip}
$a_{pm}$, mas          & 200$\pm$50     & 180$\pm$20      & 90$\pm$10         & 100$\pm$30        & 140$\pm$50     & 110$\pm$10     \\
P, years               & {\it55}        & 36.5$\pm$2      & 20.5$\pm$1        & 45$\pm$5          & 35.5$\pm$1.5   & 36$\pm$2        \\
T, year                & 1950.7$\pm$2.7 & 1994.8$\pm$1    & 1984.0$\pm$2.0    & 1971.2$\pm$1.8    & 1981.2$\pm$1.7 & 1988.9$\pm$1.8   \\
$\Omega, ^{\circ}$     & 3$\pm$20       & 32$\pm$10       & 6$\pm$61          & 326$\pm$17        & 118$\pm$9      & 296$\pm$28       \\
$\omega, ^{\circ}$     & 336$\pm$27     & 287$\pm$14      & 185$\pm$37        & 8$\pm$34          & 34$\pm$26      & 308$\pm$32       \\
e                      & 0.1$\pm$0.2    & 0.6$\pm$0.2     & 0.3$\pm$0.3       & 0.8$\pm$0.4       & 0.8$\pm$0.2    & 0.4$\pm$0.3      \\
i, $^{\circ}$          & 60$\pm$9       & 93$\pm$7        & 51$\pm$15         & 107$\pm$18        & 74$\pm$11      & 103$\pm$28       \\
  \hline
  \noalign{\smallskip}
$a_{BA}$, mas          & 830$\pm$20        & 780$\pm$30       & 460               & 520$\pm$30         & 460$\pm$30        & 350$\pm$10       \\
parallax, mas          & {\it45$\pm$0.2}   & {\it38$\pm$2}    & {\it39.2$\pm$0.4} & {\it33.2$\pm$0.3}  & {\it31$\pm$2}     & {\it19.2$\pm$0.4} \\
$\sum M, M_{\odot}$    & 2.1$\pm$0.2       & $6.4\pm0.4$      & 3.9               & $1.85\pm0.2$       & $2.5\pm0.2$       & $4.7\pm0.7$       \\
mass~of~primary, $M_{\odot}$ & {\it1.5$\pm$0.05} & {\it4.9$\pm$0.1} & {\it3.1}          & {\it1.45$\pm$0.05} & {\it1.72$\pm$0.1} & $2.4\pm0.4$      \\
mass~of~secondary, $M_{\odot}$ & $0.6\pm0.2$       & $1.5\pm0.4$      & 0.8               & $0.4\pm0.2$        & 0.8$\pm0.2$       & $2.3\pm0.4$      \\
B                      & $0.28\pm0.08$     & $0.23\pm0.1$     & 0.21              & $0.22\pm0.07$      & 0.32$\pm0.1$      & $0.49\pm0.03$     \\
$\Delta m$             & {\it$>$4$^{m}$}   & {\it$>$4$^{m}$}  & {\it$>$4$^{m}$}   & {\it$>$4$^{m}$}     & {\it$>$4$^{m}$}   & {\it1.64$^{m}\pm0.05^{m}$} \\
$\beta$                & {\it0}            & {\it0}           & {\it0}            & {\it0}             & {\it0}            & $0.18\pm0.01$      \\
$B-\beta$              & $0.28\pm0.08$     & $0.23\pm0.1$     & 0.21              & $0.22\pm0.07$      & 0.32$\pm0.1$      & $0.31\pm0.03$      \\
Spectrum~of~primary    & F7IV-V            & A2V+A2V          & A0V               & G5IV               & F7IV              & A0V                \\
Spectrum~of~secondary  & WD or RD          & pair of dwarfs   & WD or RD          & WD or RD           & WD or RD          &                    \\
V~mag~of~primary       & {\it4.1$^{m}$}    & {\it2.2$^{m}$}   & {\it2.4$^{m}$}    & {\it5.1$^{m}$}     & {\it4.6$^{m}$}    & {\it3.7$^{m}$}     \\
V~mag~of~secondary     & $>8^{m}$          & $>6^{m}$         & $>6^{m}$          & $>9^{m}$           & $>8^{m}$          & $5.3^{m}$          \\
$M_{V}$~of~primary     & $2.4^{m}$         & $0.1^{m}$        & $0.4^{m}$         & $2.7^{m}$          & $2.0^{m}$         & $0.3^{m}$          \\
$M_{V}$~of~secondary   & $>6.4^{m}$        & $>4^{m}$        & $>5^{m}$          & $>6^{m}$           & $>6^{m}$          & $1.9^{m}$          \\
  \hline
  \end{tabular}
 \]
\end{table}

\section*{Comments on individual stars}

\subsection*{$\iota$~Vir}

This star was noted by Morrison et al. (\cite[1990]{morrison}) as an outlier in
the comparison of Carlsberg and Bordeaux meridian circle results with the
FK5 prediction. It was also considered by Wielen et al. (\cite[1999]{wielen})
as an example for a long-period astrometric binary. Based
on a comparison of Hipparcos and ground-based proper motions by the
$\Delta\mu$ method they proposed a plausible individual solution:
$P\approx200$ years, $e\approx0.5$, masses of 1.64 and 0.55, $\Delta
m\approx6.9^{m}$, $a_{BA}\approx1865$ mas, $a_{pm}\approx466$ mas and
the proper motion listed in the FK6. This prediction appears to fit rather well
as seen from our orbital solution presented in Table~5
and Fig.~1.
Inclusion of observations from these many catalogues improves the orbital fit.

The discrepancies between rather young age, low metallicity and considerable velocity
(about 48 km/s with respect to the Sun)
do not allow us to establish the status of this star.
The hidden component of 0.6 solar masses fits a main-sequence star or a white dwarf.

\subsection*{$\zeta$~UMa (Mizar)}

Hereafter the components of the stellar systems are designated following Washington
Multiplicity Catalog nomenclature rules (Mason et al., \cite[2001]{wds}).

This is a known quadruple system: the pairs Aa-Ab and Ba-Bb are divided by 14 arcsec
and move linearly one with respect to other by about 1 arcsec per
latest 250 years according to Washington Double Star Catalog (WDS, Mason et al., \cite[2001]{wds}).
The known orbital motions in the close pairs are too short-period and low-amplitude to
be detected by ground-based astrometry, except by interferometry, such as by
Hummel et al. (\cite[1998]{hummel}). The periods are 0.056 and 0.481 years
and the relative semi-major axes are 10 and 33 mas. Thus, we are inclined to
believe that the observed non-linear motion of the photocentre of the
pair Aa-Ab is due to a new unseen component. It should be designated as Mizar~Aa1.
It would cause variations of the 14-arcsec distance with an amplitude about 180 mas.
But the 250-years-long series of the relative observations of Ba-Bb w.r.t. Aa-Ab has been too noisy
to detect such variations. This series should continue.

The Hipparcos observations seem to be highly disturbed by the
multiplicity: 13\% of data were rejected and the HIP Intermediate
Astrometry still remains significantly scattered. Therefore, although the
mean position from HIP is in good agreement with the ground-based results
it is not used in our processing.

The used observations and obtained results are presented in
Table~5 and Fig.~2 where `primary' and `secondary' designate the observed
close pair Aa-Ab and the new unseen component, respectively. Edge-on orientation of
the orbit means that the observable photocentre moves almost linearly.
Some systematic errors and mistakes of ground-based astrometry can produce such an effect.
Thus, the new component is questionable.
On the other hand, some elements of our orbit presented in Table~5 are
similar to the ones of the known close pair Aa-Ab:
$\omega$ is 287$^{\circ}$ versus 284$^{\circ}$, eccentricity is 0.6 versus 0.54,
but inclination is 93$^{\circ}$ versus 60.5$^{\circ}$,
$\Omega$ is 32$^{\circ}$ versus 106$^{\circ}$.

The total mass of the close pair Aa-Ab is well determined by Hummel et al. (\cite[1998]{hummel}):
$4.9\pm0.1$ solar masses.
This helps to estimate the mass of the new component: a pair of dwarfs fits best.

\subsection*{$\gamma$~UMa (Phecda)}

Spectroscopic duplicity of this star mentioned in some catalogues seems
to be a mistake: it could not been detected because of a large rotational
velocity. No component closer than 40 mas was detected in 2
speckle-interferometric observations mentioned in the Fourth Catalog of
Interferometric Measurements of Binary Stars by Hartkopf et al.
(\cite[2009]{4d}) (hereafter -- the Fourth Catalog).
This star was detected as an astrometric ($\Delta\mu$) binary
in a comparison of Hipparcos and ground-based proper motions by Wielen et
al. (\cite[1999]{wielen}). As $\iota$~Vir mentioned above a plausible
individual solution was proposed by them: $P\approx30$ years,
$e\approx0.5$, masses of 3.07 and 0.53, $\Delta m\approx9.4^{m}$,
$a_{BA}\approx518$ mas, $a_{pm}\approx760$ mas and the proper motion
listed in the FK6. This prediction appears quite reliable as seen from our
solution presented in Table~5 and Fig.~3. A red or white
dwarf fits the parameters best. It should be noted that this orbit fits
the astrometric observations best.

\subsection*{$\kappa$~Del}

This star is mentioned as G5IV+ in HIP and other catalogues. There is a K2IV
component of 12$^{m}$ changing its separation from 10 arcsec in 1851 to
45 arcsec in 2000 (Mason et al., \cite[2001]{wds}). This star is too faint to be a physical component.
This and a common proper motion component at 216 arcsec (6500 AU) could not
influence astrometric observations.

The Fe/H estimations by Feltzing and Gonzalez (\cite[2001]{feltzing}), Nordstrom et al. (\cite[2004]{gcs}),
Holmberg et al. (\cite[2009]{gcsiii}), Takeda (\cite[2007]{takeda}) and Soubiran et al. (\cite[2008]{soubiran})
are slightly discrepant although they rule out a previously declared very high metallicity status of this star.

As for $\iota$~Vir some discrepancy between rather young age and high metallicity on the one hand and
high velocity on the other make the status of the star uncertain. 
It could have acquired the high velocity during the formation of the hidden companion,
if this star is a white dwarf.
Else it is just a low-mass main-sequence star.

The orbit is shown in Fig.~4.
Periastron passage is expected within a few years after 2011.

\subsection*{20~Oph}

Abt and Levy (\cite[1976]{abt}) calculated a spectroscopic orbit with a period of about 3.5 years.
The data used by them look insufficient to calculate a reliable orbit as
pointed out by Morbey and Griffin (\cite[1987]{griffin}).
The unseen secondary has never been detected in many speckle-interferometric
observations listed in the Fourth Catalog.
The orbit is shown in Fig.~5.

The HIP gives an acceleration solution whereas HIP2 gives a stochastic solution and
quite a different parallax which fits the spectrum, magnitude and $B-V$ better.
Low metallicity contradicts a young age of this star.

\subsection*{$\mu$~Ser}

For this star HIP and HIP2 give an acceleration solution shown as 3 open diamonds
near our orbital ellipse in Fig.~6. Fabricius and Makarov (\cite[2000]{fabr})
revised the Hipparcos observations of this star and obtained a component
solution. Its parallax and $\Delta m$ are accepted by us (see Table~5).
Also, there are five speckle-interferometric observations of this pair made
in 1995-1999 and listed in the Fourth Catalog together with the revised
Hipparcos position. The results of these six observations are shown
in Fig.~6 as light asterisks. The speckle-interferometric
observations are used together with our observational material to obtain
a better orbit. As a result it is found that the speckle-interferometric
observations fit our $a_{pm}$ when $a_{BA}=350$ mas and $B-\beta=0.31$.
Thus, for this pair both the masses are calculated.
The low precision of the mass determination may explain the discrepancy
between similar masses but different luminosities.
The same does not allow us to establish the status of the secondary.
It may be a A or F dwarf, subgiant, giant or
even a pair of late-type dwarfs.

\section*{Conclusions}

We hope that our preliminary orbits may help guide the observations so
that someone may resolve the unseen components of five of the binaries 
and highlight the need for the continual monitoring of the other.
Some unusual properties of the systems such as discrepancies between age, metallicity
and velocity may relate to the presence of the unseen components.

\section*{Acknowledgements}

We are grateful to Dr.~Roger~Griffin for providing comments on some binaries,
to Dr.~Leslie~Morrison for providing comments to the Carlsberg meridian telescope results,
to Dr.~Brian~Mason for various assistance and also to anonymous reviewer for usefull comments.

This research has made use of the Centre de Donn\'{e}es
astro\-no\-mi\-ques de Strasbourg (CDS), Astronomical Data Center (ADC) at
NASA Goddard Space Flight Center, SIMBAD astronomical database and
the Washington Double Star Catalog maintained at the U.S. Naval Observatory.

This work was supported by the Russian foundation for basic
research, RFBR, projects \#08--02--00400, \#09--02--00267.

\newpage

\begin{figure}
  \includegraphics{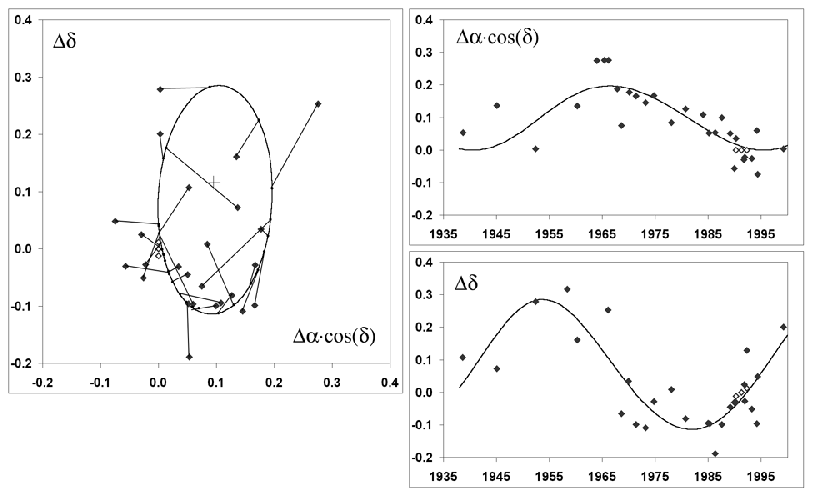}
 \caption{Photocentric orbit of $\iota$~Vir together with
 ground-based (filled diamonds) and Hipparcos (3 open diamonds) results}
 \label{ff1}
\end{figure}

\begin{figure}
 \includegraphics{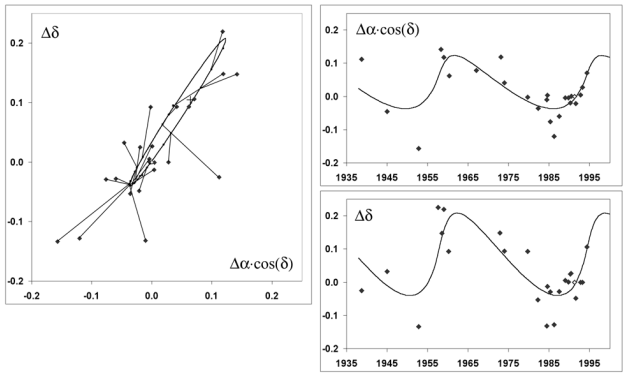}
 \caption{Photocentric orbit of $\zeta$~UMa (Mizar) together with
 ground-based (filled diamonds) and Hipparcos (1 open diamond) results}
 \label{ff2}
\end{figure}

\begin{figure}
 \includegraphics{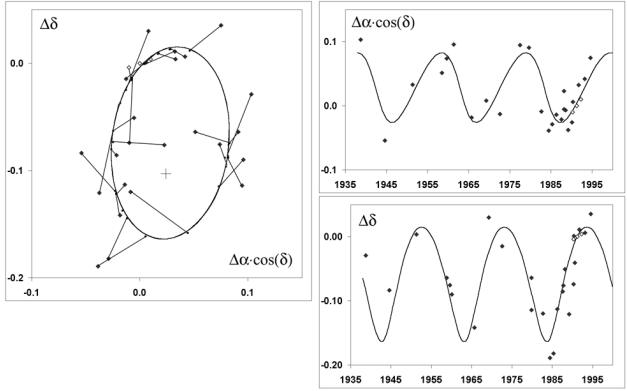}
 \caption{Photocentric orbit of $\gamma$~UMa (Phecda) together with
 ground-based (filled diamonds) and Hipparcos (3 open diamond) results}
 \label{ff3}
\end{figure}

\begin{figure}
 \includegraphics{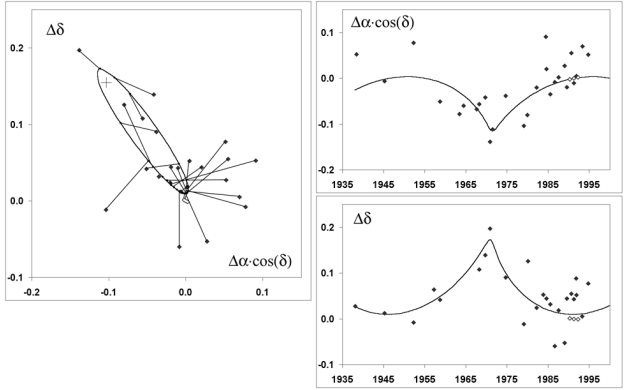}
 \caption{Photocentric orbit of $\kappa$~Del together with
 ground-based (filled diamonds) and Hipparcos (3 open diamond) results}
 \label{ff4}
\end{figure}

\begin{figure}
 \includegraphics{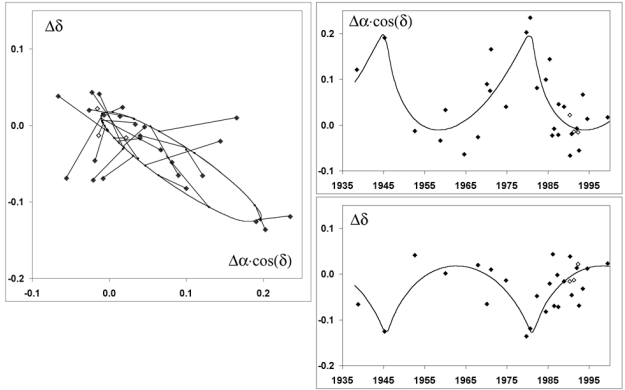}
 \caption{Photocentric orbit of 20~Oph together with
 ground-based (filled diamonds) and Hipparcos (3 open diamond) results}
 \label{ff5}
\end{figure}

\begin{figure}
 \includegraphics{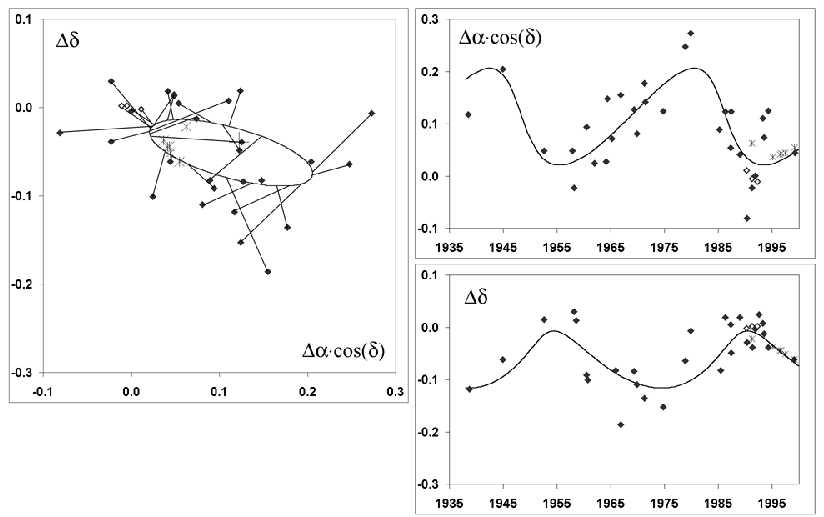}
 \caption{Photocentric orbit of $\mu$~Ser together with
 ground-based (filled diamonds), Hipparcos acceleration solution (3 open diamond),
 revised Hipparcos component solution by Fabricius and Makarov (\cite[2000]{fabr})
 with the assumption $B-\beta=0.31$ (a separate light asterisk)
 and the results of speckle-interferometric observations
 with the assumption $B-\beta=0.31$ (light asterisks)}
 \label{ff6}
\end{figure}


\begin{thebibliography}{99}

\bibitem{abt} Abt, H.A., Levy, S.G., 1976, ApJS, 30, 273.

\bibitem{abtmorrel} Abt, H.A., Morrell, N.I., 1995, ApJS, 99, 135.

\bibitem{bala} Balachandran, S., 1990, ApJ, 354, 310.

\bibitem{ccb} Chereul, E., Creze, M., Bienayme, O., 1999, A\&AS, 135, 5.

\bibitem{hip} ESA, 1997, The Hipparcos and Tycho Catalogues, ESA SP-1200.

\bibitem{fabr} Fabricius, C., Makarov, V.V., 2000, A\&AS, 144, 45.

\bibitem{feltzing} Feltzing, S., Gonzalez, G., 2001, A\&A, 367, 253.

\bibitem{fk5b} Fricke, W., Schwan, H., Lederle, T., et al., 1988, Ver\"{o}ff. Astron. Rechen-Institut Heidelberg, 32.

\bibitem{pmfs} Gontcharov, G.A., Andronova, A.A., Titov, O.A., Kornilov, E.V., 2001, A\&A, 365, 222.

\bibitem{i} Gontcharov, G.A., Kiyaeva, O.V., 2002, A\&A, 391, 647.

\bibitem{rv} Gontcharov, G.A., 2006, Astronomy Letters, 32, 759.

\bibitem{4d} Hartkopf, W.I., Mason, B.D., Wycoff, G.L., et al., Fourth Catalog of Interferometric Measurements of Binary Stars, 2009, URL: http://ad.usno.navy.mil/wds/int4.html

\bibitem{hamc} Hispano-Argentinian Meridian Catalogue (HAMC), 2001, Real Inst. y Obs. de la Armada en San Fernando, Observatorio Astronomico Felix Aguilar.

\bibitem{gcsiii} Holmberg, J., Nordstrom, B., Andersen, J., 2009, A\&A, 501, 941.

\bibitem{hummel} Hummel, C.A., Mozurkewich, D., Armstrong, J.T., et al., 1998, AJ, 116, 2536.

\bibitem{hip2} van Leeuwen F., 2007, A\&A, 474, 653.

\bibitem{wds} Mason, B., Wycoff, G.L., Hartkopf, W.I., 2001, AJ, 122, 3466, http://www.usno.navy.mil, http://www.usno.navy.mil/USNO/astrometry/optical-IR-prod/wds/wmc

\bibitem{griffin} Morbey, C.L., Griffin, R.F., 1987, ApJ, 317, 343.

\bibitem{morossi} Morossi, C., Di Marcantonio, P., Franchini, M., et al., 2002, ApJ, 577, 377.

\bibitem{morrison} Morrison, L.V., Argyle, R.W., Requieme, Y., et al., 1990, A\&A, 240, 173.

\bibitem{gcs} Nordstrom, B., Mayor, M., Andersen, J., et al., 2004, A\&A, 418, 989.

\bibitem{w2j00} Rafferty, T.J., Holdenried, E.R., 2002, The W2j00 catalog, USNO, http://ad.usno.navy.mil/proj/W2J00/

\bibitem{soubiran} Soubiran, C., Bienayme, O., Mishenina, T.V., et al., 2008, A\&A, 480, 91.

\bibitem{takeda} Takeda, Y., 2007, PASJ, 59, 335.

\bibitem{herst} Tucker, R.H., Buontempo, M.E., Gibbs, P., Swifte, R.H.D., 1983, Royal Greenwich Obs. Bull., 189.

\bibitem{wielen} Wielen, R., Dettbarn, C., Jahrei\ss, H., et al., 1999, A\&A, 346, 675.

\bibitem{fk6} Wielen, R., Schwan, H., Dettbarn, C., et al., 1999, Ver\"{o}ff. Astron. Rechen-Institut Heidelberg, 35.

\end{thebibliography}
\end{document}